\documentclass[12pt]{iopart}

\usepackage[utf8]{inputenc}
\usepackage[english]{babel}
\usepackage[T1]{fontenc}
\usepackage{graphicx}
\usepackage{bm}
\usepackage{xcolor} 
\usepackage{siunitx}
\DeclareSIUnit\pxl{pixels}
\DeclareSIUnit\ppm{ppm}
\DeclareSIUnit\ppb{ppb}
\DeclareSIUnit\ppt{ppt}

\usepackage{cleveref}

\usepackage{lipsum}

\crefname{figure}{Figure}{Figures}
\crefname{equation}{Equation}{Equations}
\crefname{table}{Table}{Tables}

\newcommand{\jk}[1]{\textcolor{black}{#1}}

\DeclareSIUnit\pxl{pixels}
\DeclareSIUnit\ppm{ppm}
\DeclareSIUnit\ppb{ppb}
\DeclareSIUnit\ppt{ppt}
\DeclareSIUnit{\liter}{$\ell$}

\graphicspath{{Figures/Julien/}}

\begin{document}
\title{Maximizing energy deposition by shaping few-cycle laser pulses}

\author{{Julien Gateau}$^1$, Alexander Patas$^2$,  Mary Matthews$^1$, Sylvain Hermelin$^{1,4}$, Albrecht Lindinger$^2$, {J\'er\^ome Kasparian}$^{1,3}$, {Jean-Pierre Wolf}$^1$}
\address{$^1$ Group of Applied Physics, Universit\'e de Gen\`eve, 22 chemin de Pinchat, 1211 Geneva 4, Switzerland} 
\address{$^2$ Institut F\"ur Experimental Physik, Freie Universit\"{a}t Berlin, Arnimallee 14, 14195 Berlin, Germany}
\address{$^3$ Institute for Environmental Sciences, Universit\'e de Gen\`eve,  Boulevard Carl-Vogt 66, CH-1211 Geneva 4, Switzerland}
\address{$^4$ University of Lyon, Université Claude Bernard Lyon 1, CNRS, Institut Lumière Matière, F-69622 Villeurbanne, France}

\ead{jerome.kasparian@unige.ch}
\vspace{10pt}
\begin{indented}
\item[]\today
\end{indented}

\begin{abstract}
We experimentally investigate the impact of pulse shape on the dynamics of laser-generated plasma in rare gases. Fast-rising triangular pulses with a slower decay lead to early ionization of the air and depose energy more efficiently than their temporally reversed counterparts. As a result, in both argon and krypton, the induced shockwave as well as the plasma luminescence are stronger.  
This is due to an earlier availability of free electrons to undergo inverse Bremsstrahlung on the pulse trailing edge. Our results illustrate the ability of adequately tailored pulse shapes to optimize the energy deposition in gas plasmas. 
\end{abstract}


\section{Introduction}

Filamentation is a nonlinear propagation regime of high-power, ultrashort lasers~\cite{ChinHLLTABKKS2005,berge2007ultrashort,couairon_femtosecond_2007}.  It conveys a typical intensity of  \jk{\SI[exponent-product = \times]{5e13}{W/cm^2}} at \SI{800}{nm}~\cite{KaspaSC2000} over distances largely exceeding the Rayleigh length, sustained by a dynamic balance between self-focusing due to the Kerr effect, and defocusing by higher-order nonlinear effects, including ionization. Due to this ionization as well as absorption by the plasma and photodissociation of molecules from the air~\cite{schwarz2000ultraviolet,petit2010production,saathoff2013laser}, thermal energy is deposited in the wake of the laser beam. It induces a shockwave~\cite{yu2003sonographic} and generates a depleted channel (or "density hole")~\cite{vidal2000modeling} with a duration in the millisecond-range~\cite{Lahav2014,jhajj2014demonstration,cheng2013effect}. This channel can be used for various atmospheric applications~\cite{Wolf2018}, to guide optical beams~\cite{jhajj2014demonstration}, drill a transmitting channel through fog~\cite{delacruz2016increased}, or trigger high-voltage discharges~\cite{zhao1995femtosecond,comtois2003triggexp,rodriguez2002triggering,point2016energy} and lightning~\cite{KaspaAAMMPRSSYMSWW2008a}.

Developing these applications however requires optimizing ionization and thermal effects in the filament. In this paper, we investigate the role of the pulse waveform in that regard. In the context of filamentation, pulse shaping~\cite{judson1992teaching} was shown to allow controling the filamentation onset position~\cite{HeckSL2006}, white-light and ionization yield~\cite{AckerSLKRSLLWBBW2006}, or the deposition of molecular rotational energy~\cite{Zahedpour2014}. Furthermore, temporal Airy pulses were used to optimize high-aspect ratio nanomachining~\cite{Goette2016}, or the poration of cell membranes~\cite{Courvoisier2016}.
Based on the shaping of extremely broadband (\SI{\ge 200}{nm}) few-cycle pulses, we investigate the influence of the pulse shape on the contributions of ionization and heating to energy deposition by ultrashort laser pulses. In particular, using asymmetric triangular shapes, we show that pulses with a sharp (few-cycle) intensity rise at their front lead to early ionization and depose energy more efficiently than slower-rising ones. Therefore, the energy deposition can be optimized by the choice of an adequate pulse shape. 

\section{Material and Methods}

\subsection{Experimental setup}
\begin{figure}
\begin{center}
\includegraphics[width=0.7\columnwidth]{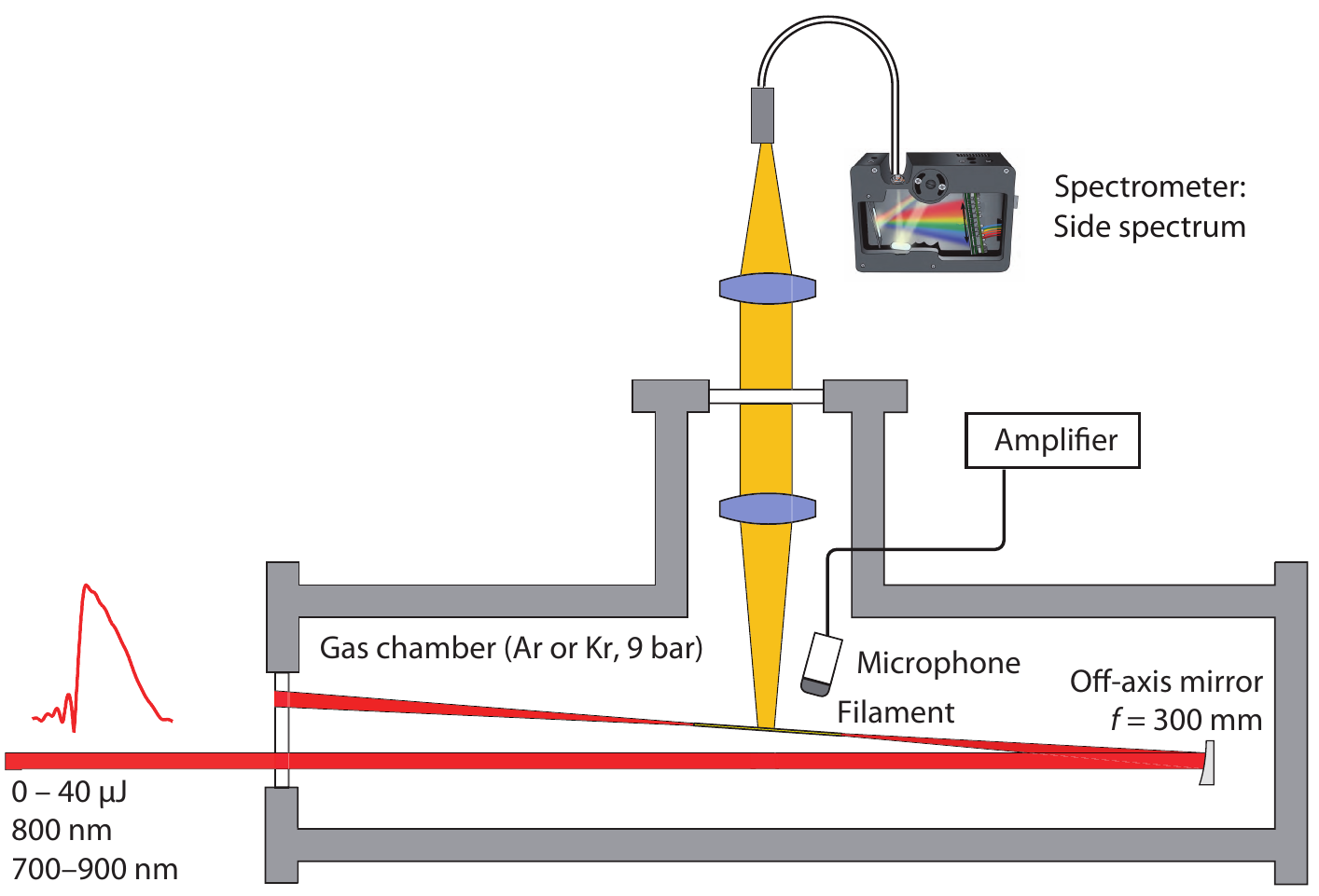}
\end{center}
\caption{Experimental setup.}
\label{Fig:setup_fil_berlin}
\end{figure}

\begin{figure}
\begin{center}
\includegraphics[width=0.7\columnwidth,keepaspectratio]{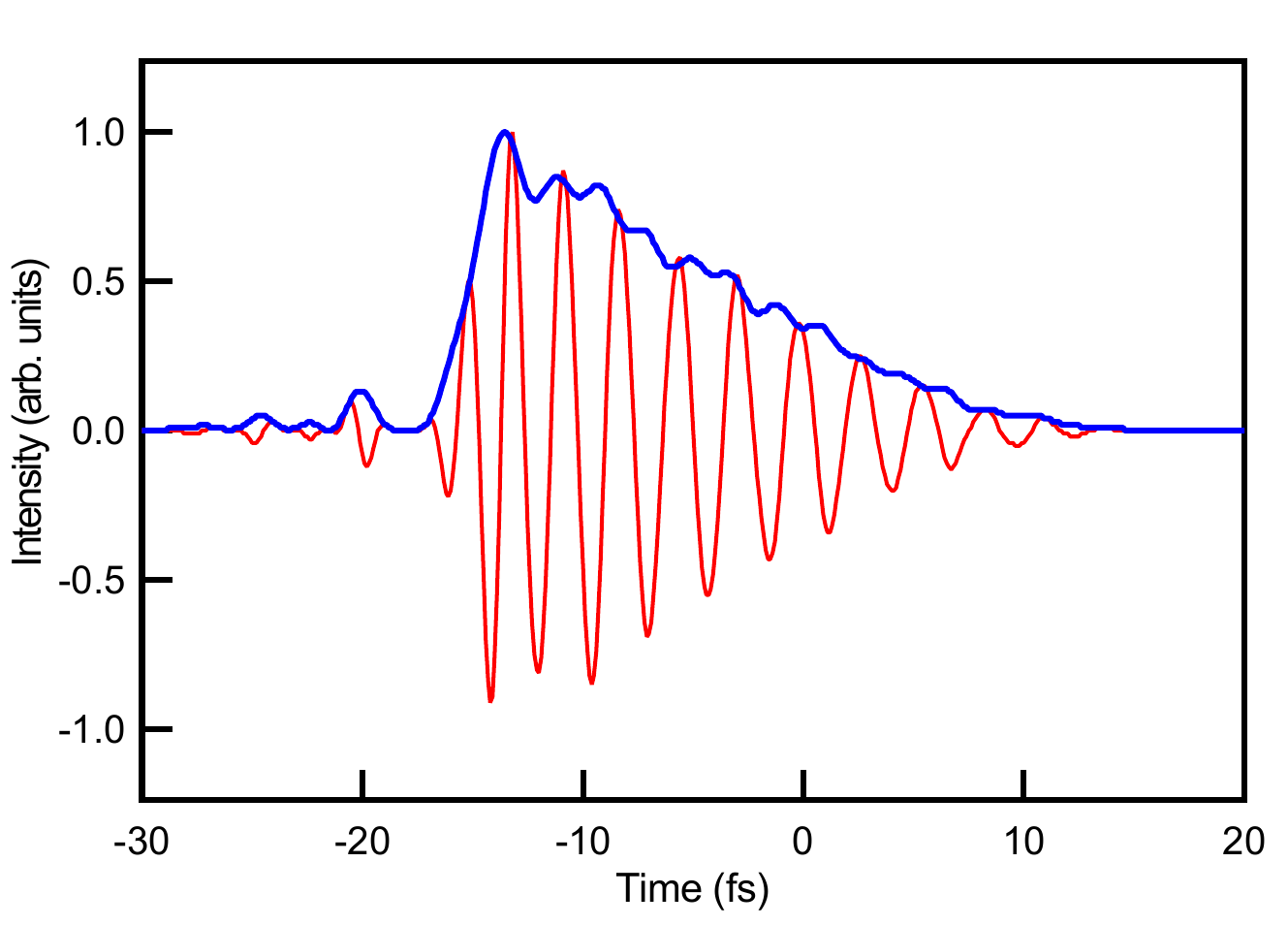}
\end{center}
\caption{Fast-rising asymmetric triangular pulse (\SI{5}{fs} rise time, \SI{30}{fs} fall time), reconstructed from the phase applied by the waveform synthetizer}
\label{Fig:exotic_PS}
\end{figure}

The  experimental setup (\cref{Fig:setup_fil_berlin}) uses a chirped-pulse amplification Ti:Sa laser chain delivering sub-\SI{40}{fs} pulses, of \SIrange{0.4}{1.4}{mJ}, centered at \SI{807}{nm}, with a bandwidth of \SI{46}{nm}, at a repetition rate of \SI{1}{kHz}.
The laser pumps two successive filamentation stages at atmospheric pressure in air, each one followed by a recompression stage based on a pair of chirped mirrors. This setup, described in detail by Hagemann \emph{et al.}~\cite{hagemann2013supercontinuum}, delivers pulses with a spectrum spanning from \SIrange{450}{1000}{nm}. It can be recompressed to \SI{5}{fs} pulse duration (two optical cycles) with a 4-$f$ zero-dispersion compressor with a dual-mask liquid crystal modulator (640~pixels), that allows simultaneous and independent shaping of the phase and the amplitude of the laser field \cite{hagemann2013supercontinuum,weiner2000femtosecond}. 
We use this waveform synthetiser to produce asymmetric triangular pulses, with either a fast-rising (\cref{Fig:exotic_PS}) or a fast-falling edge as short as 1--2 optical cycles (\SI{5}{fs}). 
The same waveform synthetizer is used to pre-compensate the dispersion of the pulse on its way to the target.
The output pulses are characterized by a transient grating FROG~\cite{trebino1997measuring,schmidt2008poor,schmidt2012optimal}. 
At the shaper output, the pulse spectral width is reduced to \SI{200}{nm}, covering the range from \SIrange{700}{900}{nm}.
 
The pulse is then launched into a gas cell (\cref{Fig:setup_fil_berlin}) filled with \SI{9}{bar} of argon or krypton, where it is focused from its initial beam diameter of \SI{0.9}{cm} at 1/$e^2$, using an $f$~=~\SI{300}{mm} off-axis gold parabolic mirror.

The gas ionization in the filaments, close to the focal point, is simultaneously characterized by two approaches. First, a fiber spectrometer (OceanOptics HR4000) collects the light emission on the side of the filaments with a resolution of \SI{0.4}{nm}. Simultaneously, an electret microphone records the acoustic shockwave generated by the plasma. Indeed, the energy $E_{p}$ deposited by the laser pulse locally heats the gas, producing an acoustic shockwave~\cite{Lahav2014,jhajj2014demonstration} that has been demonstrated to be representative of the free carrier density in the case of Fourier-limited pulses~\cite{yu2003sonographic,hosseini2004multi}. 
The relative energy of the shockwave was estimated by calculating the standard deviation of the microphone signal over a time interval of 2~ms. 

\noindent

\subsection{Numerical simulations}
Numerical simulations of the plasma dynamics were performed by adapting an air plasma model~\cite{schubert2016optimal} to rare gases.
The free electron density $N_\textrm{e}$ evolves as:

\begin{equation}
\frac{\textrm{d}N_\textrm{e}}{\textrm{d}t}  
= R_{\textrm{PPT}} + R_{\textrm{coll}} + R_{\textrm{aval}} - R_{\textrm{recomb}}  \label{eq:Ne}
\end{equation}
where
\begin{eqnarray}
R_{\textrm{PPT}} & = & N_{\textrm{At}} W(I_\textrm{L}) \label{eq:R3} \\
R_{\textrm{coll}} & = & \alpha N_e \label{eq:R2}\\ 
R_{\textrm{aval}} & = & \frac{1}{\omega_0^2 \tau^2+1} \frac{q_e^2 \tau}{c \varepsilon_0 m_e U_{\textrm{p}}} I_\textrm{L} N_e \label{eq:R1}\\ 
R_\textrm{recomb} &=& 1.138\times10^{-11} T_\textrm{e}^{-0.7} N_\textrm{e}^2  \label{eq:R4},
\end{eqnarray}
$I_\textrm{L}$ being the incident laser intensity, $N_\textrm{At}$ the density of neutral atoms, and $U_\textrm{p}$ the ionization potential. $W(I_\textrm{L})$ describes the multiphoton / tunnel ionization probability calculated with the Perelomov, Popov, Terentev (PPT) formula~\cite{perelomov1966ionization}. 
The terms explicited in Equations~(\ref{eq:R2}) to~(\ref{eq:R4}) account for collision~\cite{Rapp1965a} and avalanche ionization~\cite{papeer2014extended,papeer2015corrigendum}, and electron-ion recombination~\cite{zhao1995femtosecond}, respectively.  
$\tau = {1}/{\nu_{en}} = 1 / \left(10^{-13}  N_{\textrm{Mol}}  \sqrt{T_{e,[eV]}}\right)$ is the inverse of the electron-neutral atom collision frequency. 

The evolution of the free electron temperature $T_\textrm{e}$~\cite{fernsler1979nrl,papeer2014extended} is governed by inverse Bremsstrahlung, Joule heating, energy exchanges with the atoms and ions (at temperature $T_\textrm{g}$), the excess energy $U_\textrm{e}$ in the ionization, losses of kinetic energy due to collision and avalanche ionization as well as impact excitation as measured by Petrov \emph{et al.}~\cite{Petrov2002}: 
\begin{eqnarray}
\frac{\textrm{d}T_\textrm{e}}{\textrm{d}t}
	&=& \frac{2 J_\textrm{L}}{3 N_\textrm{e}  k_\textrm{B}} + \frac{2 q_\textrm{e} \mu_\textrm{e} E^2 }{3 k_\textrm{B}}  
	- 2 (T_\textrm{e} - T_\textrm{g})  \frac{m_\textrm{e} \nu_\textrm{c}}{M_{\textrm{gas}}} \nonumber\\ 
	&+&   \left[R_{\textrm{PPT}} U_\textrm{e}
	- \left(R_{\textrm{coll}} + R_{\textrm{aval}} \right) U_{\textrm{p}}  
	 \right] \cdot \frac{2}{3 k_\textrm{B} N_\textrm{e}} 
	-  \frac{2 N_{\textrm{Tot}}}{3 k_\textrm{B}} R_{\textrm{impact}} 
\end{eqnarray}

Here, $\mu_e$ is the electron mobility \cite{zhao1995femtosecond}:
\begin{equation}
\mu_e (\textrm{m}^2/\textrm{V}\cdot \textrm{s}) = -\frac{N_0}{3N_\textrm{Tot}}\left(\frac{5\times10^5+E_0}{1.9\times10^4+26.7\times E_0}\right)^{0.6}
\label{eq:mobility}
\end{equation}
$N_\textrm{Tot}$ being the initial density of atoms, $N_0$ the atom density at \SI{1}{atm} and ${E_0=E N_0 / N_\textrm{Tot}}$. The Bremsstrahlung rate is

\begin{equation}
J_\textrm{L} = \frac{4 \pi q_\textrm{e}^2 N_\textrm{e} \nu_\textrm{ei}} {m_e  c  \varepsilon_0 \left(\omega_0^2 + \nu_\textrm{ei}^2\right)} I_\textrm{L}.
\end{equation}

The kinetic temperature of the heavy species evolves as~\cite{papeer2014extended} 
\begin{equation}
\frac{\textrm{d}T_\textrm{g}}{\textrm{d}t}
	= 2 (T_\textrm{e} - T_\textrm{g}) \frac{m_\textrm{e} \nu_\textrm{c} N_\textrm{e}}{M_{\textrm{gas}} N_{\textrm{Tot}}} ,
\end{equation}
where the collision rate of electrons on heavy species is the sum of their collisions with atoms and positive ions~\cite{papeer2014extended}: 
\begin{equation}
\nu_{c} = 10^{-13} N_{\textrm{At}} \sqrt{T_{e,[eV]}} + 10^{-11} N_\textrm{e} T_{e,[eV]}^{-1.5}
\label{eq:nu_c}
\end{equation}

In the model, the laser pulses are described by their envelope intensity, so that coherent effects related to chirp and resonant intermediate levels in multiphoton ionization are not taken into account~\cite{Assion1996,Yakovlev1998,Lee2001,Itakura2003}. Similarly, the nonlinear propagation of the ultrashort pulses is not considered, nor the spatial aspects, including transport as well as intensity and electron density gradients. \jk{In comparison between triangular and Gaussian pulses, the latter are dimensioned to feature the same pulse energy and peak intensity as the former. This leads to a duration of \SI{16.4}{fs} FWHM for triangular pulses with \SI{5}{fs} rise time and \SI{30}{fs} fall time or vice-versa.}

\section{Results and discussion}
 
 \begin{figure}
\begin{center}
\includegraphics[width=1\columnwidth,keepaspectratio]{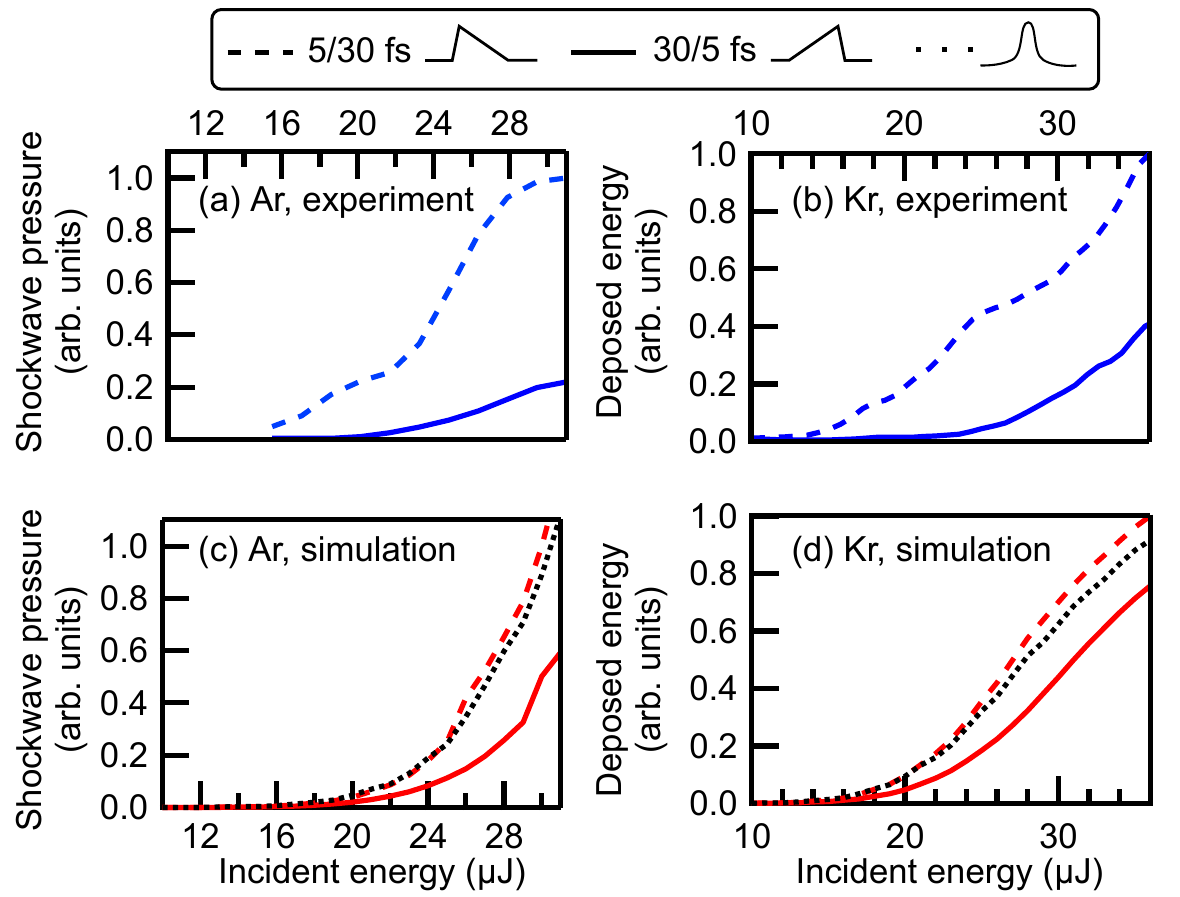}
\end{center}
\caption{(a,b) Experimental shockwave pressure and (c,d) simulated energy deposition in 9 bar (a,c) argon and (b,d) krypton by asymmetric triangular pulses with \SI{5}{fs} rise and \SI{30}{fs} fall time, or vice-versa. Simulations of a Gaussian input pulse with the same energy and peak intensity \jk{(corresponding to \SI{16.4}{fs} FWHM)} is provided for reference in Panels (c) and (d)}
\label{Fig:P9_ST_Varie_Ar}
\end{figure}\noindent

\cref{Fig:P9_ST_Varie_Ar}(a,b) compares the shockwave amplitudes induced by triangular pulses with the same duration, but fast- and slower-rising fronts, respectively.
The shockwave transient overpressure induced by the laser is proportional to the deposited energy~\cite{Lahav2014,jhajj2014demonstration}. As expected, it increases with increasing incident pulse energy. 
However it does not follow a power law as a multiphoton ionization process would predict. It even tends to start saturating above \SI{\sim 30}{\micro\J} in Ar. This beginning saturation can mainly be attributed to the depletion of neutral atoms at high energy.

The shockwave amplitude is 2--3 times higher for a triangular pulse with \SI{5}{fs} rise time and \SI{30}{fs} (dashed line), than its counterpart of the same duration but opposite fronts (\SI{30}{fs} rise, \SI{5}{fs} fall, solid line). This asymmetric behavior is more marked in argon (Fig. \ref{Fig:P9_ST_Varie_Ar}a) than in krypton (Fig. \ref{Fig:P9_ST_Varie_Ar}b). It is also observed for triangular pulses with a fast front of \SI{5}{fs} and a slow front of 10 or \SI{20}{fs}, in both argon and Krypton. 

\begin{figure}
\begin{center}
\includegraphics[width=\columnwidth,keepaspectratio]{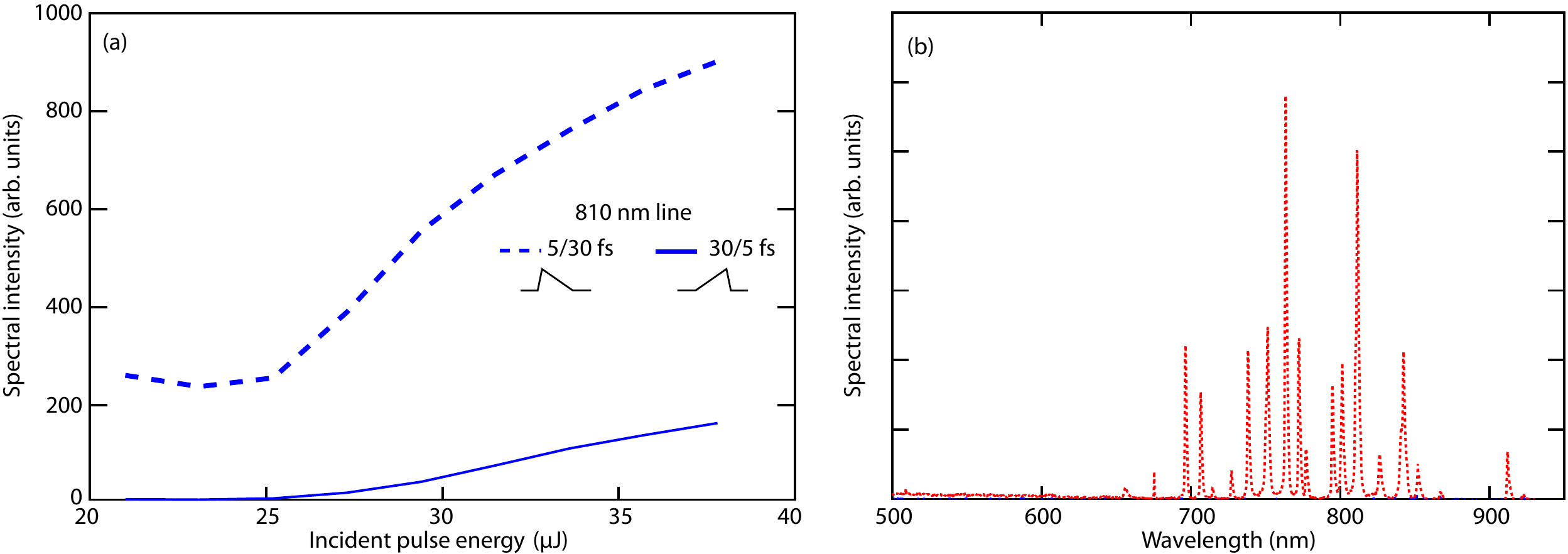}
\end{center}
\caption{\jk{(a)} Pulse energy dependence of the \SI{810}{nm} argon plasma line in \SI{9}{bar} argon, excited by asymmetric triangular pulses with \SI{5}{fs} rise and \SI{30}{fs} fall time, or vice-versa. \jk{(b)} Typical measured spectrum.}
\label{Fig:P9_SS_Ar}
\end{figure}

Similarly, the optical plasma emission measured sideways in argon, which also depends on the plasma temperature, is more efficient in the case of a fast-rising pulse as compared to a slower-rising pulse of the same duration and energy (Fig.~\ref{Fig:P9_SS_Ar}a).  \jk{Note that the plasma emission (Fig.~\ref{Fig:P9_SS_Ar}b) is independent from the pulse shape, as the excitation is non-resonant.}

 \noindent

\begin{figure}[tbh]
\begin{center}
\includegraphics[width=0.7\columnwidth,keepaspectratio]{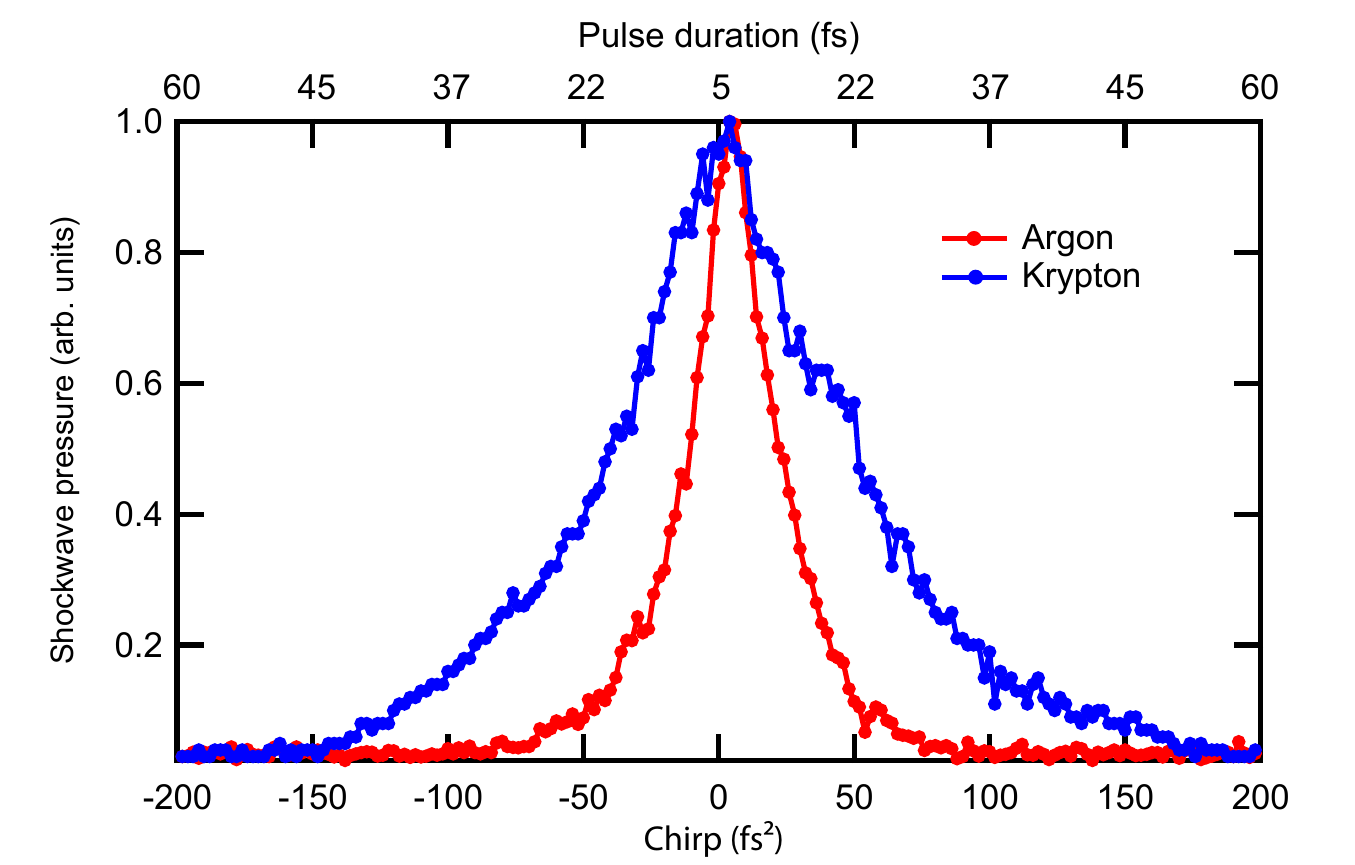}
\end{center}
\caption{Energy deposited by a \SI{47}{\micro\J} chirped Gaussian pulse as a function of the negative and positive quadratic phase, i.e chirp in krypton and argon. The pulse durations are FWHM estimated from the pulse shape on the wave synthetizer.}
\label{Fig:chirp}
\end{figure}\noindent

In order to distinguish between the role of the pulse intensity profile and resonant effects related to the carrier-phase chirp, we investigated the effect of chirp on the energy deposition efficiency. More specifically, we used the waveform synthetizer to apply various chirps to a Gaussian pulse with \SI{200}{nm} bandwidth, and measured the resulting energy deposition, in both argon and krypton.  The curves are close to symmetric (\cref{Fig:chirp}), showing that pulses of same shape, similar durations and opposite chirps depose energy with a similar efficiency. This behaviour contrasts with the chirp-asymmetry of ionization observed in the context of higher-harmonic generation~\cite{Assion1996,Yakovlev1998,Lee2001,Itakura2003}.

\begin{figure}[th]
\begin{center}
\includegraphics[width=0.75\columnwidth,keepaspectratio]{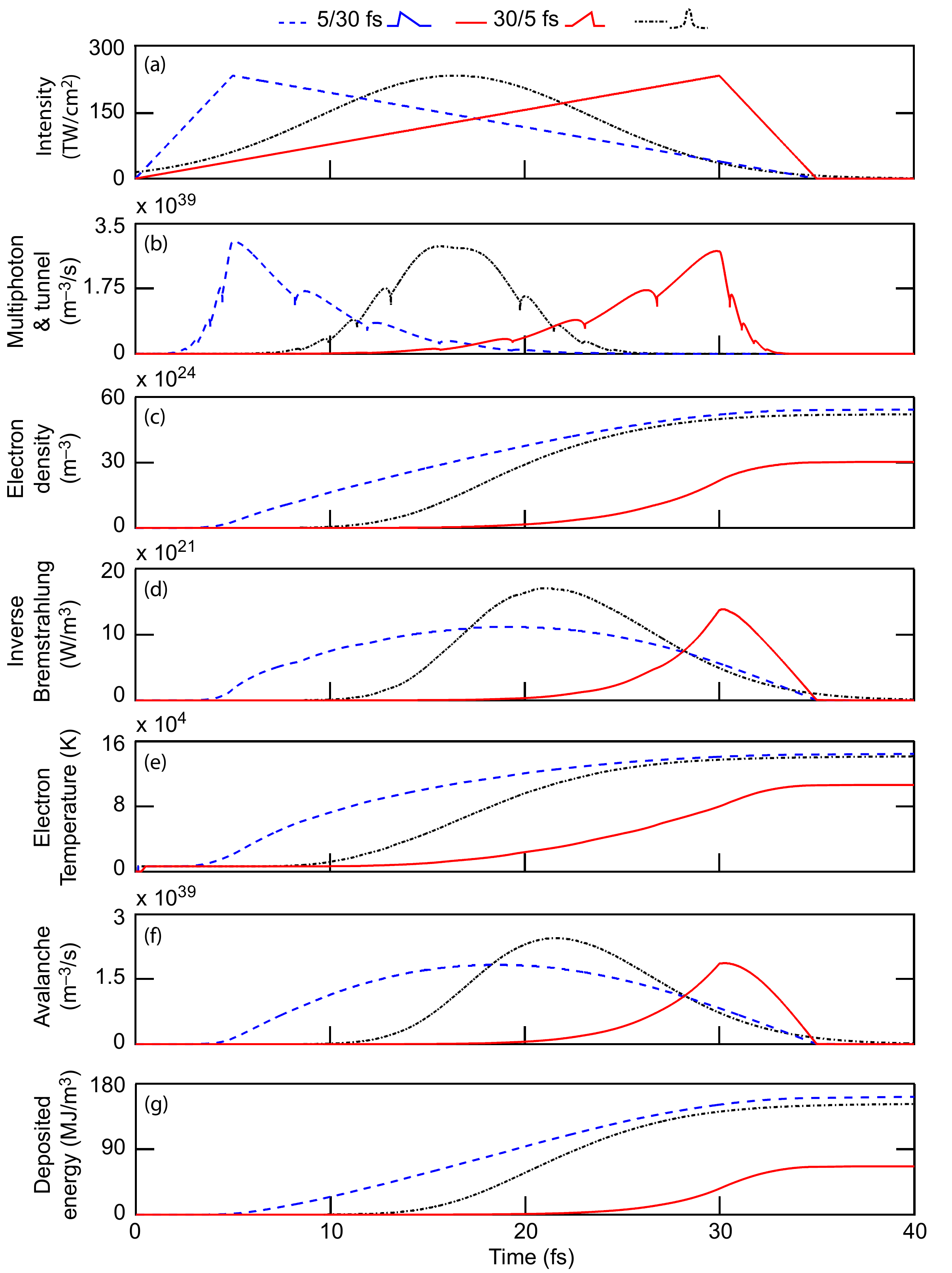}
\caption{Numerical simulation of the evolution of the plasma in argon, ionized by asymmetric triangular pulses of opposite directions and carrying \SI{20}{\micro\J} energy each. (a) Incident laser pulse intensity; (b) Multiphoton / tunnel ionization; (c) Free electron density; (d) Inverse Bremsstrahlung; (e) Free electron temperature; (f) Avalanche ionization; (g) Density of deposited energy}
\label{Fig:Density_T_Pulse_ST}
\end{center}
\end{figure}\noindent

The effect of the pulse shape on the energy deposition can be understood by considering that the laser pulses depose energy in the gas by two processes. The first one, ionization, is highly nonlinear and mostly occurs at the intensity maximum, i.e. at the beginning of fast-rising pulses and at the end of slower-rising ones. The second energy deposition process, inverse Bremsstrahlung, is linear in intensity, and proportional to the free electron density. It therefore occurs only after the ionization, i.e., on the trailing edge of the pulse. It will therefore be more efficient on slower-decaying pulses, i.e., on fast-rising ones. 

This two-step energy deposition also explains why the difference between the energy deposited by the fast-rising and slower-rising pulses is less pronounced in krypton (\cref{Fig:P9_ST_Varie_Ar}b) than in argon (\cref{Fig:P9_ST_Varie_Ar}a). Krypton has a lower ionization potential (\SI{14.00}{eV}, i.e. 9~photons at \SI{780}{nm}) as compared to argon (\SI{15.76}{eV}, 10~photons). As a consequence, the ionization is slightly less nonlinear, so that the ionization efficiency depends less critically on the intensity. Ionization therefore occurs over a slightly longer time interval around the peak of the pulse, limiting the effect of the asymmetric pulse shape. This effect also explains the fact that the energy deposed in krypton depends less on the incident pulse duration than that in argon (see the broader curve in \cref{Fig:chirp}).

This interpretation is supported by our numerical simulations. As examplified in  \cref{Fig:Density_T_Pulse_ST}, multiphoton / tunnel ionization (Panel b) starts close to the intensity peak of the pulse (Panel a). The rise in electron density (panel (c) is followed shortly by inverse Bremsstrahlung heating (Panel d) of the electrons. The higher electron temperature (Panel e) allows avalanche ionization to switch on (Panel e) for the remaining duration of the pulse, further contributing the free electron density (Panel c). As a result, the sharp rising pulse releases more, hotter, free electrons, hence deposes substantially more energy (Panel g). 
The simulations reproduce well the asymmetry of the measured deposited energy as characterized via the shockwave overpressure, as well as its dependence on the input energy (\cref{Fig:P9_ST_Varie_Ar}(c,d)). Simultaneously, the simulated electron densities approach the initial density of the neutral gas atoms ($N_\textrm{At}=$~\jk{\SI[exponent-product = \times]{2.5e25}{m^{-3}}}), confirming that  the saturation of the signal in argon is mainly due to the depletion of the latter.

It is remarkable that this fairly good agreement is obtained with simulations based on the pulse intensity shape only, without the need to consider the temporal phase within the pulse, nor the resonant processes in the atom ionization. This confirms that in our experiment, the energy deposition is indeed governed by the pulse shape rather than resonant processes. 
The observed effects are quite robust with regard to the pulse shape, and could be effective in perturbed conditions like the open atmosphere.

Its benefit is further demonstrated by comparing the energy deposition by a fast-rising triangular pulse with a Gaussian one featuring the same peak intensity and pulse energy \jk{(\SI{16.4}{fs} FWHM)}. As shown in \cref{Fig:P9_ST_Varie_Ar}(c,d), the early ionization in the fast-rising triangular pulses maximizes the energy deposition. The dynamics detailed in \cref{Fig:Density_T_Pulse_ST} confirms this description, showing in particular the later occurrence of Bremsstrahlung and avalanche ionization.

\subsection{Conclusion}

In conclusion, we highlight the impact of the pulse shape on the energy deposition by laser filaments in rare gases. Due to an earlier ionization within the pulse duration, fast-rising pulses heats the plasma more efficiently, allowing avalanche ionization. This result demonstrates the relevance of pulse shaping with sharp profiles for optimizing remote ionization and energy deposition in filaments. It therefore  provides an opportunity to control the plasma dynamics by adequate pulse shaping.

\textbf{Acknowledgments}. 
This work was supported by the ERC advanced grant "Filatmo" and the SNF NCCR MUST grant. M.M. acknowledges funding from MHV fellowship grant number: PMPDP2-145444 and NCCR MUST Womens Postdoc Award. S.H. acknowledges cofunding under FP7-Marie Skłodowska-Curie -- NCCR MUST
program (200021-117810). \jk{A. L. acknowledges the Klaus Tschira Foundation (KTS) for financial support (project 00.314.2017).}
We gratefully acknowledge fruitful discussions with L. W\"oste, theoretical support from N. Berti and E. Schubert, as well as technical support by M. Moret.

\bibliography{mabiblio}
\bibliographystyle{iopart-num}
\end{document}